\documentclass{article}

\usepackage{microtype}
\usepackage{graphicx}
\usepackage{subfigure}
\usepackage{booktabs} 
\usepackage{CJKutf8}
\usepackage{hyperref}
\usepackage{url}

\def\VersionPublic{1}  
\def\VersionSubmit{2}  
\def\submission{\VersionPublic}  
\if\submission\VersionSubmit
    \usepackage{icml2024}
\else
    \usepackage[accepted]{icml2024}
\fi

\usepackage{amsmath}
\usepackage{amssymb}
\usepackage{mathtools}
\usepackage{amsthm}

\usepackage[capitalize,noabbrev]{cleveref}
\usepackage{longtable}

\theoremstyle{plain}

\theoremstyle{definition}

\theoremstyle{remark}

\usepackage[textsize=tiny]{todonotes}

\usepackage{listings}
\usepackage{wrapfig}
\usepackage{floatflt}
\usepackage{multirow}
\usepackage{float}
\usepackage{multicol}
\usepackage{enumitem}  

\usepackage{tabularx} 

\usepackage{xcolor}  
\usepackage{colortbl}  

\definecolor{orange}{RGB}{221, 124, 79}
\definecolor{purple}{RGB}{108, 97, 175}
\definecolor{green}{RGB}{98, 156, 53}
\definecolor{paleorchid}{RGB}{183, 118, 175}
\definecolor{gray}{rgb}{0.5,0.5,0.5}
\definecolor{darkerblue}{rgb}{0,0.08,0.45}
\definecolor{darkergreen}{RGB}{21, 152, 56}
\definecolor{darkerred}{RGB}{220, 35, 120}
\definecolor{gray94}{gray}{.94}
\definecolor{gray90}{gray}{.90}
\definecolor{gray85}{gray}{.85}

\newcommand{\ul}{\underline}

\newcommand{\orange}[1]{\textcolor{orange}{#1}}
\newcommand{\purple}[1]{\textcolor{purple}{#1}}
\newcommand{\green}[1]{\textcolor{green}{#1}}
\newcommand{\paleorchid}[1]{\textcolor{paleorchid}{#1}}

\newcommand{\grow}[1]{\rowcolor{gray94}{#1}}

\usepackage{pifont}%
\usepackage{xspace}
\usepackage{placeins}
\if\submission\VersionSubmit
    \icmltitlerunning{Submission and Formatting Instructions for ICML 2024}
\else
    \icmltitlerunning{VQDNA: Unleashing the Power of Vector Quantization for Multi-Species Genomic Sequence Modeling}
\fi

\begin{document}

\twocolumn[
\icmltitle{VQDNA: Unleashing the Power of Vector Quantization for Multi-Species Genomic Sequence Modeling}



\icmlsetsymbol{equal}{*}


\begin{icmlauthorlist}
\icmlauthor{~~Siyuan Li}{equal,west,zju}
\icmlauthor{~~Zedong Wang}{equal,west}
\icmlauthor{~~Zicheng Liu}{west,zju}
\icmlauthor{~~Di Wu}{west,zju}
\icmlauthor{~~Cheng Tan}{west,zju}
\icmlauthor{~~Jiangbin Zheng}{west,zju}
\icmlauthor{~~Yufei Huang}{west,zju}
\icmlauthor{~~Stan Z. Li$^\dag$}{west}
\end{icmlauthorlist}


\icmlaffiliation{west}{AI Lab, Research Center for Industries of the Future, Westlake University, Hangzhou, 310024, China}
\icmlaffiliation{zju}{College of Computer Science and Technology, Zhejiang University, Hangzhou, 310058, China}

\icmlcorrespondingauthor{Stan Z. Li}{stan.z.li@westlake.edu.cn}

\icmlkeywords{Vector quantization, Self-supervised learning, Pre-training}

\vskip 0.3in
]



\printAffiliationsAndNotice{\icmlEqualContribution} 

\begin{abstract}

Similar to natural language models, pre-trained genome language models are proposed to capture the underlying intricacies within genomes with unsupervised sequence modeling. They have become essential tools for researchers and practitioners in biology.
However, the \textit{hand-crafted} tokenization policies used in these models may not encode the most discriminative patterns from the limited vocabulary of genomic data.
In this paper, we introduce VQDNA, a general-purpose framework that renovates genome tokenization from the perspective of genome vocabulary learning. By leveraging vector-quantized codebook as \textit{learnable} vocabulary, VQDNA can adaptively tokenize genomes into \textit{pattern-aware} embeddings in an end-to-end manner.
To further push its limits, we propose Hierarchical Residual Quantization (HRQ), where varying scales of codebooks are designed in a hierarchy to enrich the genome vocabulary in a coarse-to-fine manner.
Extensive experiments on 32 genome datasets demonstrate VQDNA's superiority and favorable parameter efficiency compared to existing genome language models. Notably, empirical analysis of SARS-CoV-2 mutations reveals the fine-grained pattern awareness and biological significance of learned HRQ vocabulary, highlighting its untapped potential for broader applications in genomics.

\end{abstract}

\section{Introduction}
\label{sec:intro}

Genomics, which refers to the study of genomes-the complete set of DNA instructions within an organism, enables scientists to delve into the molecular machinery of life~\citep{yang2011gcta, encode2020expanded}.
It provides critical insights into genetic coding and expression that orchestrate the development, functioning, and reproduction of living organisms, thereby prompting a paradigm shift in biological discovery, unlocking mysteries of multifactorial traits, genetic diseases, and evolution~\citep{locke2015genetic, visscher201710, andersson2020determinants}. 
By leveraging deep learning techniques, breakthroughs in genomics have burst onto the scene, showcasing their preeminence in addressing a broad spectrum of biological applications, such as splicing regulation and gene expression prediction~\citep{Kelley2015BassetLT, Zhou2015PredictingEO, Avsec2021EffectiveGE}, DNA methylation prediction~\citep{Vidaki2017DNAMF, Angermueller2017DeepCpGAP}, chromatin accessibility~\citep{Min2017ChromatinAP}, promoter prediction~\citep{Lai2019iProEP, Le2022BertPromoter} and more.

\begin{figure}[t]
    \vspace{-0.5em}
    \begin{center}
    \hspace{-1.4em}
    \includegraphics[width=1.05\linewidth]{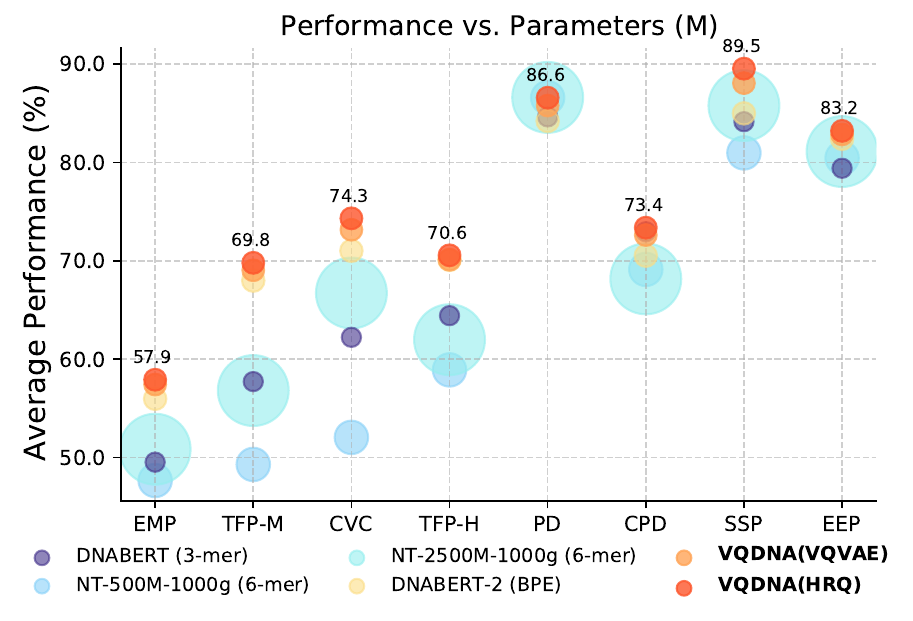}
    \end{center}
    \vspace{-1.75em}
    \caption{Performance of fine-tuned VQDNA and other genome language models across downstream tasks on 32 datasets, including Epigenetic Mark Prediction (EMP) for Yeast, Transcription Factor Prediction on mouse and human genome (TFP-M and TFP-H), Covid Variants Classification (CVC), Promoter Detection (PD), Core Promoter Detection (CPD), Splice Site Prediction (SSP), and Editing Efficiency Prediction (EEP). The circle size indicates the parameter scale of each model. Notably, NT-2500M-1000g is with 2537M model parameters, while our VQDNA has only 110M.}
    \label{fig:gue_metric_param}
    \vspace{-1.25em}
\end{figure}

In parallel, large-scale genomic data has been readily accumulated, which presents the opportunity, differing from specialized advancements, for digging out generalizable patterns that can be directly fine-tuned for various downstream tasks.
Drawing inspiration from the success of natural language models~\citep{Devlin2019Bert, Brown2020GPT, Ouyang2022Training}, genome language models have been introduced by representing genomes as languages for unsupervised genomic sequence modeling.
DNABERT~\citep{Ji2021DNABert} first explores the language model-style pre-training on the human genome.
Nucleotide Transformers~\citep{Dalla2023NT} is pre-trained on massive multi-species genomes, increasing cross-species diversity.
HyenaDNA~\citep{Nguyen2023HyenaDNA} targets the unique but challenging extra-long sequence issue and strikes remarkable accuracy-efficiency trade-offs.
Very recently, DNABERT-2~\citep{Zhou2023DNABert2} pioneered the use of Byte Pair Encoding (BPE)~\citep{Sennrich2015NeuralMT} to iteratively merge the co-occurring nucleotides that might be relevant in genomics.

Along this line, tokenization has become an integral part of genome language models, significantly influencing the model's perception and interpretation of genomes \citep{Zhou2023DNABert2}. 
The commonly used k-mer combines adjacent sets of k-length nucleotide bases through a sliding window of specified strides.
BPE, however, iteratively merges the statistically most co-occurring segments regardless of the topological distance.
Although more precise tokenization strategies have been introduced, these \textit{hand-crafted} methods may not represent sufficient information from the limited vocabulary (A, T, C, and G, four nucleotide bases) of genomes and thus cannot guarantee the derived word embeddings encoding the most discriminative genomic patterns~\citep{Chelba2017Ngram}. Thus, merging genome segments solely according to \textit{hand-crafted} policies might mislead the training into subpar representations, resulting in inevitable sample inefficiency and non-generalizability.
In this paper, we argue that if we can derive a \textit{learnable} genome vocabulary that records the most discriminative patterns from input genomes, we can thus use it as an off-the-shelf weapon to tokenize genomes into pattern-aware embeddings for subsequent pre-training.

To this end, we reconstruct the genome tokenization into a \textbf{discriminative genome vocabulary learning} problem and propose VQDNA, a novel framework eschewing \textit{hand-crafted} schemes and instead relying entirely on the VQ-VAE~\citep{NIPS2017VQVAE} tokenizer, which computes \textit{pattern-aware} embeddings with the VQ codebook as \textit{online-optimizable} genome vocabulary.
Built upon this concept, we further conjecture that the limited original vocabulary of genomes may conceivably hamper discriminative codebook learning, resulting in the loss of fine-grained details trapped in the four nucleotides.
To further push the limits of VQ tokenizer, we present Hierarchical Residual Quantization (HRQ), where varying scales of codebooks are designed in a hierarchical structure with coarse-grained semantics concentrated in the lower layers, and fine-grained details in the higher layers to expand the vocabulary for perceptually rich codebook learning in a coarse-to-fine progressive manner.

We comprehensively evaluate the effectiveness of VQDNA on GUE benchmark~\citep{Zhou2023DNABert2} with 28 datasets and 4 additional genome datasets as illustrated in Figure~\ref{fig:gue_metric_param} and Sec.~\ref{sec:comp} involving the input sequence lengths from 63 up to 32k.
%
To further validate our methods on the unique but meaningful extra-long sequence issue, we extend the input length of VQDNA (HRQ) to a maximum of 32k, allowing fair comparisons with HyenaDNA in Species Classification (SC) tasks.
Extensive experiments show that our VQDNA, as a general-purpose framework for multi-species genomic sequence modeling, can handle large and diverse genome analysis tasks and hits state-of-the-art across 32 datasets of varying input lengths while striking favorable complexity-accuracy trade-offs.
More importantly, empirical analysis of the SARS-CoV-2 demonstrates the \textbf{fine-grained pattern-awareness }and\textbf{ biological significance} of HRQ vocabulary, revealing its potential for broader applications in biology.


Our contributions can thus be summarized as follows:
\begin{itemize}[leftmargin=1.15em]
\vspace{-0.5em}
    \item 
    We push the boundaries of genome tokenization from the fresh perspective of genome vocabulary learning, presenting the VQDNA framework to learn a VQ codebook as discriminative genome vocabulary for pattern-aware genome language tokenization in an end-to-end manner.
    \vspace{-0.25em}
    \item 
    An HRQ tokenizer is designed to progressively enrich the originally limited genome vocabulary with a hierarchy of varying scales of codebooks in a coarse-to-fine manner. This hierarchical design delivers performance on par with the state-of-the-art models while using fewer parameters.
    \vspace{-1.25em}
    \item 
    Extensive experiments across 32 datasets verify the exceptional generalizability of VQDNA. Empirical study on SARS-CoV-2 mutations shows the biological significance and potential of VQDNA among existing models.
\end{itemize}

\section{Related Work}
\label{sec:related}

\subsection{Pre-trained Genome Language Models}
Genomics has witnessed rapid advances in recent decades thanks to the emergence of new technologies that facilitate high-throughput DNA sequencing.
This precipitous drop in the cost and time has led to an explosion of genomic data.
The Human Genome Project and the 1000 Genomes Project~\citep{Byrska20221KGP} have successfully sequenced thousands of individual genomes, identifying millions of genetic variants.
In addition to the human genome, genomes from other organisms have also been extensively sequenced and analyzed.
The abundance of data provides the opportunity to explore pre-trained language models in genomics that can be adapted to various downstream tasks.

DNABERT~\citep{Ji2021DNABert} introduces the first pre-trained genome language model based on BERT~\citep{Devlin2019Bert} architecture. 
They pre-train the BERT Transformer solely on human genome to develop a general understanding of DNA and then fine-tune the models on task-specific datasets, including Eukaryotic Promoter Database for promoters, ENCODE ChIP-seq for TF binding sites, and more.
Techniques are adjusted to suit the DNA characteristics, such as the masking scheme and next-sentence prediction.
Similar to natural language models, tokenization is critical in the model's perception and interpretation of genomes. They utilize overlapping k-mer to incorporate contextual information from genomes in tokenization. Despite its shortcomings, DNABERT has inspired almost all the subsequent genome language models as a ground-breaker.
Nucleotide Transformer~\citep{Dalla2023NT} proposes a new family of transformer-based genome language models.
These models, ranging from 500M up to 2.5B parameters, are pre-trained on the human reference genome, 3,202 genetically diverse human genomes, and 850 multi-species genomes. It is a huge leap in terms of the volume and diversity of training data, directly leading to superior performance.
As for tokenization, they first use non-overlapping k-mer instead of the overlapping version in DNABERT. They empirically show that tokenizing DNA into different mers exerts quite diverse performance which highlights the value of genome tokenization. 
Moreover, empirical analyses like attention maps confirm the learned representations can reconstruct human genetic variants and distinguish between key genomic elements like exons, promoters, and enhancers.

The newly emerged DNABERT-2~\citep{Zhou2023DNABert2} systematically discusses current genome tokenization techniques and first adapts SentencePiece~\citep{Kudo2018Sentencepiece} with BPE to tokenize genome sequences. 
They also integrate Attention with Linear Biases (ALiBi)~\citep{Press2021ALiBi}, FlashAttention~\citep{Dao2022FlashAttention}, LoRA fine-tuning \citep{Hu2021LoRA} and more practical techniques to overcome the architectural limitations of existing genome language models.
Additionally, several genomic benchmarks are published~\cite{fishman2023gena, Nguyen2023HyenaDNA}. MUSE~\cite{iclr2024bend} and Genome Understanding Evaluation (GUE)~\cite{Zhou2023DNABert2} provide multi-species genome analysis with well-calibrated data separation, task setting, and evaluation metrics, resolving the lack of standard benchmarks for existing genome language models.


\subsection{Vector Quantization}
First pioneered in image compression~\citep{Gray1984VQ}, vector quantization (VQ) as a parametric method has demonstrated tremendous success in generating high-fidelity patterns by discretizing latent space. 
Holistically, VQ quantizes the continuous latent from the encoder into discrete vectors by replacing them with the closest embeddings from the \textit{learnable} codebook.
VQ-VAE~\citep{NIPS2017VQVAE} as the cornerstone first introduces a vector-quantized learning framework comprising training and generation phases. It encodes image pixels into latent features and then searches for the nearest token to each corresponding feature vector. The image is thereby reconstructed through the decoder. During training, an annealing procedure is employed to guide the quantization, helping avoid posterior collapse issues softly. 
%
%
Thereafter, a multi-scale variant~\citep{NIPS2019VQVAE2} is proposed. \citet{Dhariwal2020jukebox} adds random restart policy to avoid codebook collapse.
VQGAN~\citep{CVPR2021VQGAN} leverages GPT-2 as the generator and employs adversarial loss and feature-level perceptual losses in the training stage, which shows improved reconstruction quality over VQ-VAE.
As such, VQGAN-based variants have been adapted to video and more scenarios~\citep{Yu2023MAGViT, Yu2023LFQ}.
MaskGIT~\citep{Chang2022MaskGITMG} proposes a new paradigm where masked tokens are predicted by attending to tokens from all directions. 
%
RQ~\citep{Lee2022RVQ} refines the latent feature by quantized residuals, and \citet{Huh2023StraighteningOT} examines critical challenges in VQ training.
MAGE~\cite{cvpr2023mage} predicts randomly masked VQ tokens in the latent space~\cite{Li2023MIMSurvey} that first combine both self-supervised pre-training~\cite{he2021masked, li2024genurl} and image generation into one framework.
LQAE~\citep{Liu2023LQAE} tokenizes the input into lexical representations with frozen BERT word embeddings.
Recently, FSQ~\citep{Mentzer2023FSQ} boosts the quantization efficiency with finite-scalar implicit codebook. VQ techniques are also adopted in AI4S methods~\cite{iclr2024SaProt, wu2024mape, wu2024psc} to model both sequential and structural information.

As yet, the quantized posterior has proven effective in unleashing the full expressivity of complex multi-modal distributions such as images and videos. To the best of our knowledge, however, there is no such attempt to leverage VQ for genome language models.
In the following sections, we will first incorporate VQ into genome tokenization in our proposed three-stage VQDNA framework. Next, we describe HRQ vocabulary learning architecture and discuss its advantages with extensive experimental results.

\begin{figure*}[t]
    \centering
    \includegraphics[width=1.0\linewidth]{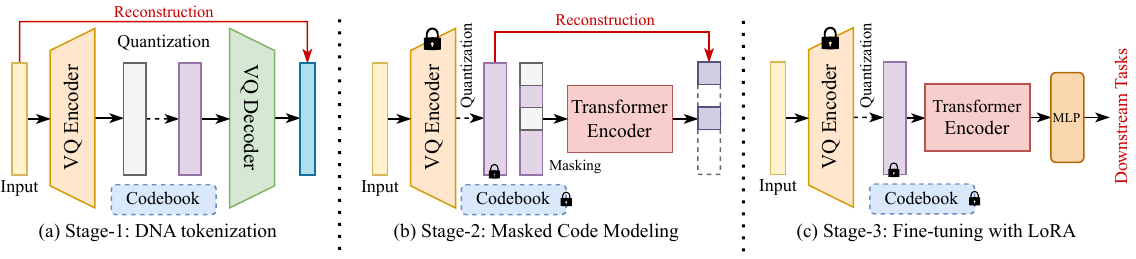}
\vspace{-2.5em}
    \caption{An overview of our three-stage training pipeline of VQDNA. (a) VQ genome vocabulary learning with large-scale multi-species genome sequences. (b) Masked modeling pre-training of the Transformer encoder with frozen genome vocabulary. (c) Fine-tuning the pre-trained encoder with an MLP head for various downstream genome analysis tasks.
    }
    \label{fig:pipeline}
    \vspace{-0.75em}
\end{figure*}

\section{Methodology}
\label{sec:prelim}
This paper aims to develop a general-purpose framework by leveraging VQ codebook as \textit{learnable} genome vocabulary that can adaptively tokenize inputs into \textit{pattern-aware} word embeddings for genomic sequence modeling to serve multiple downstream tasks.
The core idea behind this is to learn a discriminative genome vocabulary consisting of discrete code embeddings that can then get assigned to corresponding latent features via a nearest-neighbor lookup for genome tokenization. 
By optimizing this vocabulary to minimize quantization objectives, the codebook embeddings can essentially represent a dictionary of pattern-aware data clusters learned in a completely self-supervised paradigm.

In this section, we introduce our three-stage VQDNA training framework, as shown in Figure~\ref{fig:pipeline}, to multi-species genomic sequence modeling. We first propose to incorporate the renowned VQ-VAE~\cite{NIPS2017VQVAE} instead of \textit{hand-crafted} methods into tokenization for \textit{pattern-aware} genome vocabulary learning, as illustrated in Sec.~\ref{subsec:VQDNA-VAE}. 
Moreover, we further conjecture that the limited vocabulary of genome sequences may conceivably hamper discriminative codebook learning, resulting in the loss of fine-grained patterns trapped in the original four nucleotides. To tackle this problem, we propose hierarchical residual quantization (HRQ) in Sec.~\ref{subsec:HRQ}, to progressively enrich the genome vocabulary with a hierarchy of varying scales of codebooks in a coarse-to-fine manner.
In Sec.~\ref{subsec:Architecture}, we describe the implementation details for VQDNA vocabulary learning.

\subsection{Vector-Quantized Genome Vocabulary Learning}
\label{subsec:VQDNA-VAE}
As aforementioned, we first parameterize the tokenization as a genome vocabulary learning problem and follow the VQ-VAE to take sequence reconstruction as pre-training objectives to concurrently optimize the codebook and the VQ encoder. We present this as the base version of VQDNA.

Given the input genome sequence $X\in \mathbb{R}^{L\times d}$, an encoder $E_{\theta}(\cdot)$ with parameters $\theta$ maps $X$ into the latent space as $Z = E_{\theta}(X)\in \mathbb{R}^{L\times D}$.
With a finite vocabulary of $K$ key-value pairs as the VQ codebook, $\mathcal{C}=\{(k, e(k))\}_{k\in [K]}$, where each code (index) $k$ owns its \textit{learnable} code embedding vector  $e(k)\in \mathbb{R}^{D}$, the representation $Z$ can be quantized by the element-wise code mapping function $\mathcal{Q}(\cdot,\cdot)$:
\begin{equation}
    \vspace{-0.5em}
    M_i = \mathcal{Q}(Z_i; \mathcal{C}) = \textrm{argmin}_{k\in [K]}\|Z_i - e(k)\|_{2},
    \label{eq:codemap}
    \vspace{0.2em}
\end{equation}
where $1\leq{i}\leq{L}$, $M\in [K]^L$ denotes code mapping indices. Thus, the latent $Z_i$ can be indexed and quantized into discrete genome embeddings by the distance-wise closest 1-of-K embedding vectors within codebook $\mathcal{C}$ with assigned code $M_i$ as $\hat{Z_i} = e(M_i)$.
The decoder $G_{\phi}(\cdot)$ with parameters $\phi$ then maps the quantized embedding $\hat{Z}$ back to the input genome sequence space to reconstruct $\hat{X}$:
\begin{equation}
    \vspace{-0.5em}
    \hat{X} = G_{\phi}(\hat{Z}) = G_{\phi}(e(M)),
    \label{eq:decoder}
    \vspace{0.07em}
\end{equation}
As differentiation through the quantization is ill-posed, the straight-through-estimator (STE)~\citep{Bengio2013Estimating} is employed as gradient approximation during backward computation. 
To optimize the overall framework, the overarching models aim to minimize the VQ-VAE loss
$\mathcal{L}_\mathrm{VQ}$:
\begin{equation}
\mathcal{L}_\mathrm{VQ}=\underbrace{\mathcal{L}_{CE}(X,\hat{X})}_{\mathcal{L}_\mathrm{rec}} +  \underbrace{\| \mathrm{sg}[{Z}] - \hat{Z}\|_2^2}_{\mathcal{L}_\mathrm{code}} + \beta \underbrace{\| Z - \mathrm{sg}[\hat{Z}]\|_2^2}_{\mathcal{L}_\mathrm{commit}},
    \label{eq:vqvaeloss}
\end{equation}
where $\mathrm{sg}[\cdot]$ refers to the aforementioned stop-gradient operator, and $\beta \in [0,1]$ is a trade-off hyper-parameter (default to 0.5). 
Notably, the first term $\mathcal{L}_\mathrm{rec}$ denotes the reconstruction loss to optimize the encoder and decoder in VQ-VAE vocabulary learning (Stage-1 in Figure~\ref{fig:pipeline}). The middle term $\mathcal{L}_\mathrm{code}$ takes a squared error as the codebook loss to update code embeddings by pushing embedding vectors toward the encoder outputs. The third term $\mathcal{L}_\mathrm{commit}$ is a commitment loss, which ensures the training stability of code mapping $\mathcal{Q}(\cdot,\cdot)$.  
In this paper, we optimize the codebook $\mathcal{C}$ with the exponential moving average (EMA) of embeddings instead of the loss $\mathcal{L}_\mathrm{code}$:
\begin{equation}
    \label{eq:EMAcode}
    \hat{Z}_{i} = (1-\alpha) Z_{i} + \alpha \hat{Z}_{i},
\end{equation}
where $\alpha$ is the momentum coefficient. The EMA update of the codebook in Eq.~(\ref{eq:EMAcode}) can reduce the training instability caused by updating conflicts of the certain code from latent tokens of different subjects \citep{NIPS2019VQVAE2}. 

After obtaining the learned codebook $\mathcal{C}$, we can reuse it as an off-the-shelf genome vocabulary to tokenize genome sequences into \textit{pattern-aware} genome embeddings for language model pre-training.
Subsequently, we can store the tokenized data for stage-2 pre-training (described in Appendix~\ref{app:implement}) and conduct the same masked pre-training and downstream fine-tuning as DNABERT-2 for our VQDNA with the trained VQ-VAE vocabulary (described in Sec.~\ref{sec:exp}).


\begin{figure*}[t]
\centering
\begin{minipage}{0.67\linewidth}
    \includegraphics[width=1.0\linewidth]{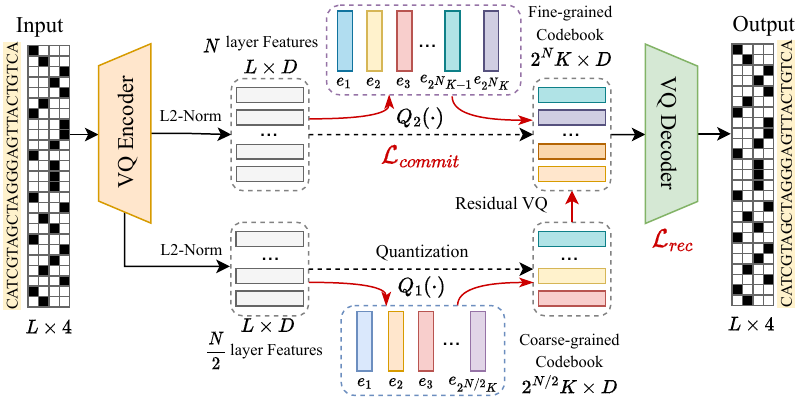}
    \vspace{-2.5em}
    \caption{Illustration of our Hierarchical Residual Quantization (HRQ) as genome word embedding for VQDNA framework.
    We instantiate HRQ with a 6-layer encoder and decoder with two hierarchical codebooks after the output of $3$-th and $6$-th layers in practice.}
    \label{fig:VQDNA_HRQ}
\end{minipage}
\begin{minipage}{0.32\linewidth}
    \vspace{-0.5em}
    \begin{table}[H]
    \setlength{\tabcolsep}{0.5mm}
    \centering
    \label{tab:ablation_tokenizer}
\resizebox{\linewidth}{!}{
    \begin{tabular}{ll|ccc}
    \toprule
Method         & Tokenizer   & Usage & Lin.       & FT         \\ \hline
DNABERT        & 6-mer       & 47    & 23.54      & 55.50      \\
NT-2500M-1000g & 6-mer (non) & 47    & 23.54      & 66.73      \\
HyenaDNA       & one-hot     & 100   & 5.47       & 54.10      \\
DNABERT-2      & BPE (6-mer) & 99    & 36.53      & 71.02      \\
\bf{VQDNA}     & \bf{VQVAE}  & 100   & \ul{44.76} & \ul{73.16} \\
\bf{VQDNA}     & \bf{HRQ}    & 100   & \bf{48.87} & \bf{74.32} \\
    \bottomrule
    \end{tabular}
    }
    \vspace{-0.75em}
    \caption{Analysis of tokenization efficiency. We report the tokenizer types, token usage (\%), macro F1-score (\%) of linear probing (Lin.), and fully fine-tuning (FT) for the Covid Variants Classification task, which is illustrated in Sec.~\ref{sec:ablation}. Note that 6-mer (non) utilizes non-overlapping 6-mer tokenization, and BPE (6-mer) iteratively merges the most concurrent codes in 6-mer tokenization. The token usage is the percentage of the used token in total for each tokenizer.
    }
\end{table}

    \vspace{-1.0em}
\end{minipage}
\vspace{-0.5em}
\end{figure*}

Now, we have reformulated genome tokenization from the perspective of VQ-VAE genome vocabulary learning with apparent benefits:
\textbf{(i)} The sequence nature of genomic data comfortably fits the VQ computations. Nucleotide base pairs within genomes can not only form localized motifs like promoter elements but also modulate global chromatin states, which is much akin to that of image pixels in vision, where VQ has already established its dominance. The quantized posterior has proven effective in compressing intricate multi-modal distributions, making it well-primed to encode the most discriminative genomic patterns unshackled by \textit{hand-crafted} rules and biases.
\textbf{(ii)} Genomic context plays a vital role in genome analysis tasks. Contrary to existing tokenization methods that solely concern better merging intra-sequence nucleotides, VQ tokenizer naturally records the genomic context by incorporating whole inputs into its codebook optimization implicitly rather than just regarding the intra-sequence dependencies. Empirical analysis in Sec.~\ref{sec:analysis} demonstrates both the intra- and inter-class pattern-awareness of VQ. 
The rest of this section expands on HRQ to further push the limits of genome vocabulary learning.



\subsection{Hierarchical Residual Quantization}
\label{subsec:HRQ}
Although the VQ-VAE tokenizer can provide tangible benefits above, it expands its power primarily by enlarging the codebook size.
To take a further step, however, simply expanding the codebook size is inefficient due to the codebook collapse problem, and more importantly, it may not be compatible with the nature of genomic data. Genomic data is essentially sequences consisting of four potential nucleotide bases, A, T, C, and G, at each site, which means the original vocabulary of genomes is much restricted compared to that of other modalities, such as images and natural languages. 
Through the lens of VQ tokenizer, such a limited vocabulary space might be too coarse-grained to present sufficient details for perceptually rich codebook learning. Therefore, we argue that it is necessary to design a specified protocol to disentangle such underlying intricacies within the restricted nucleotides for discriminative genome vocabulary learning.

Motivated by the success of multi-scale perception~\citep{wang2020HRNet} in visual recognition, it is appealing that we can also transfer this success from computer vision to genomics, \textit{i.e.}, to build varying scales of codebooks as multi-grained genome vocabulary and then tokenize different layers of inputs with corresponding vocabulary, which can be hierarchically aligned via residual techniques~\citep{Lee2022RVQ}.
To achieve this, we propose Hierarchical Residual Quantization (HRQ), where a hierarchy of codebooks is designed to expand the genome vocabulary in a coarse-to-fine manner.

As shown in Figure~\ref{fig:VQDNA_HRQ}, the multiple scales of codebooks are designed in a hierarchical architecture with coarse-grained semantics concentrated in the lower layers and fine-grained details in the higher layers. 
Quantization is performed sequentially from encoder layer $1$ to $N$. 
Given the hierarchical input $H^{(n)}\in \mathbb{R}^{L\times D}$ out of encoder layer $n$, a corresponding $2^n\cdot{K}$-size codebook $\mathcal{C}^{(n)}=\{(k^{(n)}, e(k^{(n)}))\}_{k\in [2^nK]}$ with each code embedding vector $e(k^{(n)})\in \mathbb{R}^{D}$ is defined. Thus, each representation $H^{(n)}$ is quantized by the same code mapping operator $\mathcal{Q}(\cdot,\cdot)$ in Eq.~(\ref{eq:codemap}):
\begin{equation}
\begin{aligned}
    \vspace{-0.5em}
    M^{(n)}_i &= \mathcal{Q}(H^{(n)}_i; \mathcal{C}^{(n)})\\
              &= \textrm{argmin}_{k\in [2^nK]}\|H^{(n)}_i - e(k^{(n)})\|_{2},
    \label{eq:hrqcodemap}
\end{aligned}
\end{equation}
where $1\leq{i}\leq{L}$, ${M^{(n)}}\in [2^nK]^L$ indicates the HRQ code mapping indices of ${H^{(n)}}$. As such, we derive a hierarchy of codebooks with varying perceptual granularities for hierarchical genome tokenization in a coarse-to-fine manner. With assigned $M^{(n)}_i$, the latent features of layer $n$ can be quantized as ${\hat{H}^{(n)}_i} = e(M^{(n)}_i)$. 
However, one remaining challenge is that, given the output $Z^{(n)}\in \mathbb{R}^{L\times D}$ from encoder layer $n$, how to associate the hierarchical input $H^{(n)}$ in Eq.~(\ref{eq:hrqcodemap}) with $Z^{(n)}$ to form a unified HRQ architecture.

\begin{figure}[t]
    \begin{center}
    \includegraphics[width=1.01\linewidth]{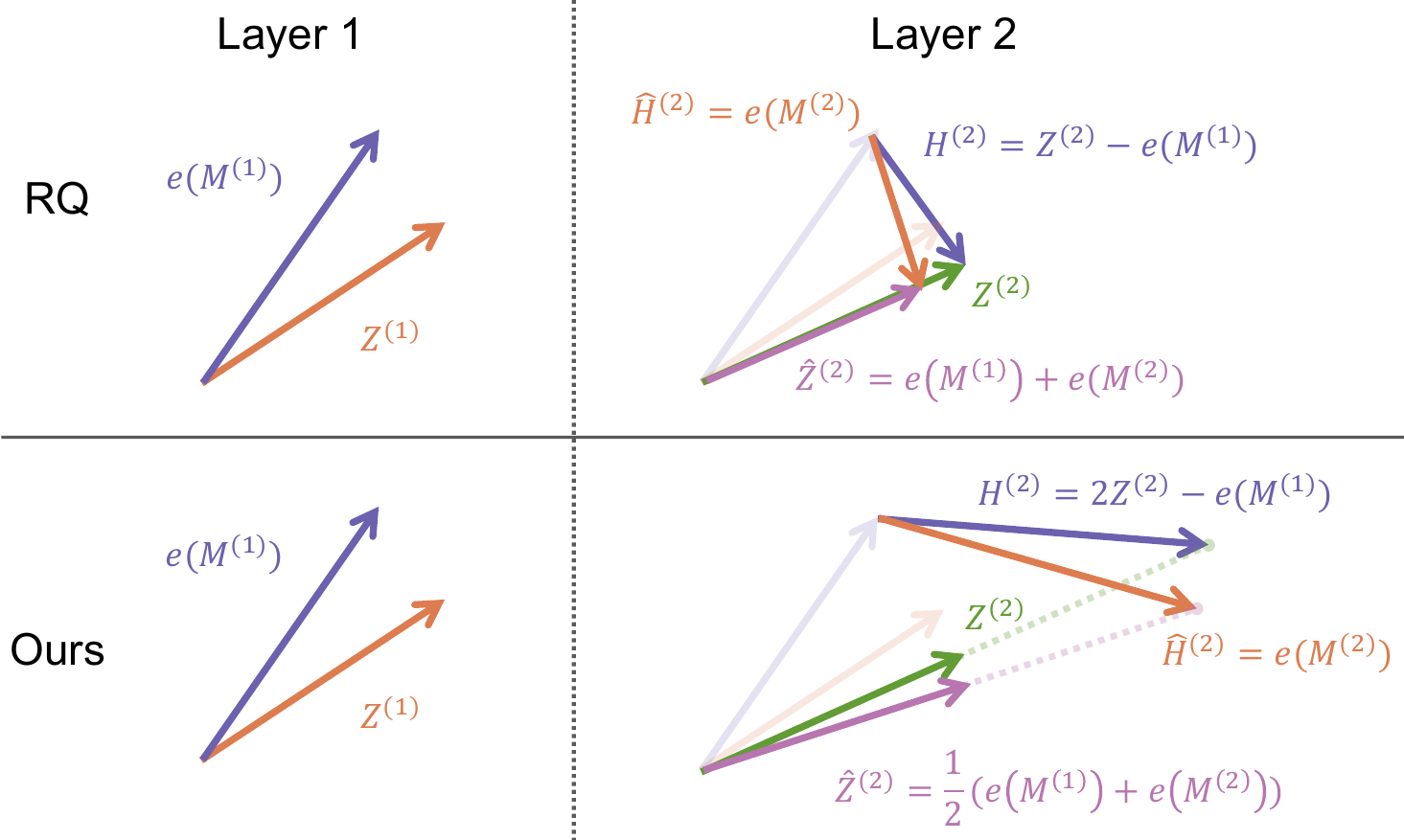}
    \hspace{-1.4em}
    \end{center}
    \vspace{-2.25em}
    \caption{Illustration of RQ and our HRQ in a two-dimensional space with a two-layer quantization case. We use \purple{purple for the current hierarchical input $H^{(n)}$(or the residual in RQ),} \green{green for the second layer encoder output $Z^{(2)}$,} \orange{orange for the input $Z^{(1)}$ and output hierarchical embeddings $\hat{H}^{(n)}$ per layer,} and \paleorchid{pale orchid for the ultimate embeddings $\hat{Z}^{(n)}$ after $n$-layer quantization.}
    }
    \label{fig:RQ_vs_HRQ}
    \vspace{-0.5em}
\end{figure}

Although \citet{Lee2022RVQ} first introduces residual quantization (RQ) to harness the training of multiple codebooks, their method is essentially designed for recursive quantization with a single input, which has not addressed the above issue of multiple inputs. 
To resolve this problem, we define a strategy to associate $H^{(n)}_i$ with $Z^{(n)}_i$ formalized as:
\begin{equation}
    \hspace{-0.25em}
    \vspace{-0.25em}
    H^{(n)}_i =
    \begin{cases*}
    2Z^{(n)}_i - e(M^{(n-1)}_i) & \text{for } $n = 2, \cdots, N$, \\
    e(M^{(1)}_i) & \text{otherwise} \label{eq:hrqhiercode}
    \end{cases*}
\end{equation}
where $1\leq{n}\leq{N}$, and $1\leq{i}\leq{L}$. Starting with the initial quantization $H^{(1)}_i=e(M^{(1)}_i)$, our HRQ calculates the code mapping $M^{(n)}$ in Eq.~(\ref{eq:hrqcodemap}), which together with $Z^{(n+1)}_i$ yields the hierarchical input $H^{(n+1)}_i$ for next layer quantization.
The motivation behind this is to resolve the alignment between $H^{(n)}_i$ and $Z^{(n)}_i$ while maintaining the scale consistency across HRQ layers, as this property has proven essential in ensuring better utilization of multiple codebooks~\citep{yu2023SPAE}.
As shown in Figure~\ref{fig:RQ_vs_HRQ}, we compare our proposed strategy with renowned RQ. Intuitively, representations computed by doubled inputs residual exhibit more favorable scale consistency across the layers.

Along this line, we obtain a hierarchy of learned codebooks. We can thereby use them as an off-the-shelf genome vocabulary to tokenize input genomes into a collection of hierarchical embeddings after $N$ layers of quantization $\mathcal{HRQ(\cdot,\cdot,\cdot)}$:
\begin{equation}
\mathcal{HRQ}(Z_i,\mathcal{C},N) = (\hat{H}^{(1)}_i,\cdots,\hat{H}^{(N)}_i),
\end{equation}
where each  ${\hat{H}^{(n)}_i} = e(M^{(n)}_i)\in \mathbb{R}^{L\times D}$ denotes the quantized genome embedding at layer $n$. As illustrated in Figure~\ref{fig:RQ_vs_HRQ}, we define the ultimate output embeddings of HRQ as 
$\hat{Z_i} = \frac{1}{N}(\sum_{n=1}^{N}\hat{H}^{(n)}_i)$, which sums the hierarchical embeddings $\hat{H}^{(n)}$ from all $N$ quantization layers in average for scale consistency.   
With the exponentially growing codebooks, the HRQ vocabulary can progressively capture the most discriminative coarse-grained semantics and the fine-grained details within input genome sequences for discriminative tokenization and subsequent masked pre-training.

\paragraph{Training of HRQ}
The overall learning objective of our proposed HRQ is defined as follows:
\vspace{-0.5em}
\begin{equation}
    \mathcal{L}_\mathcal{HRQ} =\underbrace{\mathcal{L}_{CE}(X,\hat{X})}_{\mathcal{L}_\mathrm{rec}} + \beta \underbrace{\sum_{n=1}^{N}\| Z^{(n)} - \mathrm{sg}[\hat{Z}^{(n)}]\|_2^2}_{\mathcal  {L}_\mathrm{commit}},
    \label{eq:hrqloss}
    \vspace{-0.5em}
\end{equation}
where $\beta>0$ is the same hyper-parameter as in Eq.~(\ref{eq:vqvaeloss}), and the first term is the reconstruction loss ${\mathcal{L}_\mathrm{rec}}$.
We also employ the widely-used EMA of the clustered embeddings to update codebook $\mathcal{C}$ instead of the codebook loss $\mathcal{L}_\mathrm{code}$ in Eq.~(\ref{eq:vqvaeloss}). 
The commitment loss ${\mathcal{L}_\mathrm{commit}}$ in Eq.~(\ref{eq:hrqloss}) is defined as the sum of squared errors from each layer $n$, which is different from VQ-VAE. It aims to make the quantized embeddings $\hat{Z}^{(n)}$ progressively reduce the squared error as $n$ increases. 
In such a way, HRQ disentangles the underlying semantics in the limited genome vocabulary for perceptually rich codebook learning in a hierarchically coarse-to-fine manner. 
Empirical studies in Sec.~\ref{sec:analysis} and Appendix~\ref{app:covid_analysis} demonstrate the fine-grained pattern-awareness of the HRQ vocabulary.

\begin{table}[b]
    \vspace{-2.0em}
    \setlength{\tabcolsep}{0.5mm}
    \centering
    \caption{Average performance ranking, tokenizer types, model parameters and FLOPs, and pre-training tokens on 32 genome downstream tasks.}
    \vspace{0.15em}
    \label{tab:overall_gue}
\resizebox{\linewidth}{!}{
    \begin{tabular}{ll|cccc|c}
    \toprule
Method          & Date          & Tokenizer      & \# Params. & FLOPs   & Train & Average    \\
                &               &                & (M)        & (G)     & (B)   & Rank       \\ \hline
DNABERT         & BioInfo'2021  & 3-mer          & 86         & 3.3     & 122   & 5          \\
NT-500M         & biorxiv'2023  & 6-mer          & 480        & 3.2     & 50    & 6          \\
NT-2500M        & biorxiv'2023  & 6-mer          & 2537       & 19.4    & 300   & 4          \\
DNABERT-2       & ICLR'2024     & BPE            & 117        & 1.0     & 262   & 3          \\
\grow\bf{VQDNA} & \textbf{Ours} & \textbf{VQVAE} & 86+16      & 1.1+0.5 & 262   & \ul{2}     \\
\grow\bf{VQDNA} & \textbf{Ours} & \textbf{HRQ}   & 86+17      & 1.1+0.6 & 262   & \textbf{1} \\
    \bottomrule
    \end{tabular}
    }
\end{table}

\begin{table*}[t]
    \vspace{-0.25em}
    \setlength{\tabcolsep}{2.0mm}
    \centering
    \caption{MCC (in \%) performance of Promoter Detection (PD), Core Promoter Detection (CPD), and Transcription Factor Prediction (TFP) tasks fine-tuned on GUE benchmarks.}
    \vspace{0.15em}
    \label{tab:gue_pd_tfp}
\resizebox{0.95\linewidth}{!}{
    \begin{tabular}{l|ccc|ccc|ccccc}
    \toprule
\multirow{2}{*}{Method} & \multicolumn{3}{c|}{PD}                          & \multicolumn{3}{c|}{CPD}                         & \multicolumn{5}{c}{TFP (Human)}                                                    \\ \cline{2-12} 
                        & all            & notata         & tata           & all            & notata         & tata           & 0              & 1              & 2              & 3              & 4              \\ \hline
DNABERT (3-mer)         & 90.44          & 93.61          & 69.83          & 70.92          & 69.82          & 78.15          & 67.95          & 70.90          & 60.51          & 53.03          & 69.76          \\
NT-500M-1000g (6-mer)   & 89.76          & 91.75          & 78.23          & 66.70          & 67.17          & 73.52          & 63.64          & 70.17          & 52.73          & 45.24          & 62.82          \\
NT-2500M-1000g (6-mer)  & 90.95          & 93.07          & 75.80          & 67.39          & 67.46          & 69.66          & 66.31          & 68.30          & 58.70          & 49.08          & 67.59          \\
DNABERT-2 (BPE)         & 86.77          & 94.27          & 71.59          & 69.37          & 68.04          & 74.17          & 71.99          & 76.06          & 66.52          & 58.54          & 77.43          \\
\grow\bf{VQDNA}         & \ul{90.20}     & \ul{94.05}     & {\ul 73.08}    & \ul{70.36}     & \ul{69.87}     & \ul{77.63}     & \ul{72.04}     & \ul{75.89}     & \ul{66.69}     & \ul{58.31}     & \ul{77.63}     \\
\grow\bf{VQDNA (HRQ)}   & \textbf{90.75} & \textbf{94.48} & \textbf{74.52} & \textbf{71.02} & \textbf{70.58} & \textbf{78.50} & \textbf{72.48} & \textbf{76.43} & \textbf{66.85} & \textbf{58.92} & \textbf{78.10} \\
    \bottomrule
    \end{tabular}
    }
    \vspace{-0.25em}
\end{table*}

\begin{table*}[t]
    \vspace{-0.5em}
    \setlength{\tabcolsep}{2.mm}
    \centering
    \caption{Performance of Transcription Factor Prediction (TFP), Covid Variants Classification (CVC), Splice Site Prediction (SSP), and Editing Efficiency Prediction (EEP) tasks. TFP and SSP use MCC (\%), while CVS and EEP report F1 (\%) and MCC (\%).}
    \vspace{0.15em}
    \label{tab:gue_tfp_others}
\resizebox{0.94\linewidth}{!}{
    \begin{tabular}{l|ccccc|c|c|ccc}
    \toprule
\multirow{2}{*}{Method} & \multicolumn{5}{c|}{TFP (Mouse)}                                                   & CVC            & SSP            & \multicolumn{3}{c}{EEP (gRNA)}                   \\ \cline{2-11} 
                        & 0              & 1              & 2              & 3              & 4              & Covid          & Reconstruction & K562           & Jurkat         & H1             \\ \hline
DNABERT (3-mer)         & 42.31          & 79.10          & 69.90          & 55.40          & 41.97          & 62.23          & 84.14          & 88.63          & 86.89          & 62.72          \\
NT-500M-1000g (6-mer)   & 39.26          & 75.49          & 64.70          & 33.07          & 34.01          & 52.06          & 80.97          & 90.58          & 88.94          & 63.80          \\
NT-2500M-1000g (6-mer)  & 48.31          & 80.02          & 70.14          & 42.25          & 43.40          & 66.73          & 85.78          & 90.90          & 89.34          & 66.87          \\
DNABERT-2 (BPE)         & 56.76          & 84.77          & 79.32          & 66.47          & 52.66          & 71.02          & 84.99          & 91.02          & 89.27          & 66.91          \\
\grow\bf{VQDNA}         & \ul{57.52}     & \ul{85.36}     & \ul{79.78}     & \ul{68.45}     & \ul{54.10}     & \ul{73.16}     & \ul{88.06}     & \ul{91.16}     & \ul{89.83}     & \ul{67.56}     \\
\grow\bf{VQDNA (HRQ)}   & \textbf{58.34} & \textbf{85.81} & \textbf{80.39} & \textbf{69.72} & \textbf{54.73} & \textbf{74.32} & \textbf{89.53} & \textbf{91.53} & \textbf{90.12} & \textbf{67.98} \\
    \bottomrule
    \end{tabular}
    }
    \vspace{-0.25em}
\end{table*}

\begin{table*}[t]
    \centering
    \caption{MCC (in \%) performance of Epigenetic Marks Prediction tasks with different datasets fine-tuned on GUE benchmarks.}
    \vspace{0.15em}
    \label{tab:gue_emp}
    \setlength{\tabcolsep}{1.6mm}
\resizebox{1.0\linewidth}{!}{
    \begin{tabular}{l|cccccccccc}
    \toprule
\multirow{2}{*}{Method} & \multicolumn{10}{c}{Epigenetic Marks Prediction}                                                                                                                        \\ \cline{2-11} 
                        & H3             & H3K14ac        & H3K36me3       & H3K4me1        & H3K4me2        & H3K4me3        & H3K79me3       & H3K9ac         & H4             & H4ac           \\ \hline
DNABERT (3-mer)         & 74.15          & 42.07          & 48.49          & 42.95          & 31.34          & 28.92          & 60.12          & 50.48          & 78.27          & 38.60          \\
NT-500M-1000g (6-mer)   & 72.52          & 39.37          & 45.58          & 40.45          & 31.05          & 26.16          & 59.33          & 49.29          & 76.29          & 36.79          \\
NT-2500M-1000g (6-mer)  & 74.61          & 44.08          & 50.86          & 43.10          & 30.28          & 30.87          & 61.20          & 52.36          & 79.76          & 41.46          \\
DNABERT-2 (BPE)         & 78.27          & 52.57          & 56.88          & 50.52          & 31.13          & 36.27          & 67.39          & 55.63          & 80.71          & 50.43          \\
\grow\bf{VQDNA}         & \ul{78.56}     & \ul{53.93}     & \ul{60.62}     & \ul{52.84}     & \ul{33.73}     & \ul{38.49}     & \ul{68.15}     & \ul{56.28}     & \ul{81.32}     & \ul{50.33}     \\
\grow\bf{VQDNA (HRQ)}   & \textbf{79.21} & \textbf{54.46} & \textbf{61.75} & \textbf{53.28} & \textbf{34.05} & \textbf{39.10} & \textbf{68.47} & \textbf{56.63} & \textbf{81.84} & \textbf{50.69} \\
    \bottomrule
    \end{tabular}
    }
    \vspace{-0.5em}
\end{table*}

\subsection{Implementation Details}
\label{subsec:Architecture}
We adopt the network architecture of ConvNeXt variants \cite{cvpr2022convnext, iclr2024moganet} for our tokenizers, which have Transformer-like macro designs but are more efficient. The encoder network for VQVAE and HRQ consists of a stem module and 6 residual blocks, \textit{i.e.}, $N$=6, and $D$=384. The stem projects the input data (one-hot encoded) to 256 dimensions by a 1D convolution layer with a kernel size of 5 and a stride of 1, followed by a LayerNorm~\cite{2016layernorm} and GELU activation. Each residual block contains a 1D depth-wise convolution layer (the kernel size of 7) and 2 full-connected layers to form the inverted bottleneck~\cite{cvpr2018mobilenetv2} (expanding 4 times). The architecture of the de-tokenizer (the decoder of the VQDNA tokenizer) is symmetrical to the tokenizer in Figure~\ref{fig:VQDNA_HRQ}, except for using 1D de-convolution layers instead. The output sequence length of the tokenizer is the same as the input.
The standard VQVAE vocabulary learning uses a codebook size of 512, while the HRQ version uses the size of 384. In practice, we instantiate the HRQ decoder only with the $3$-th and $6$-th layers codebooks. The masked pre-training and downstream adaptation details are described in Sec.~\ref{sec:exp}.


\section{Experiments}
\label{sec:exp}
\subsection{Experimental Setup}
In pre-training stages, we follow the pre-training recipes in DNABERT-2~\citep{Zhou2023DNABert2} that pre-training the VQ tokenizer and BERT-Base Transformer encoder on the human genome~\citep{Ji2021DNABert} with 2.75B nucleotide bases and the multi-species genome~\citep{Zhou2023DNABert2} with 32.49B nucleotide bases. In the pre-training stage-1, VQDNA variants are pre-trained one epoch by AdamW~\citep{iclr2019AdamW} optimizer with a batch size of 1024 and a basic learning rate of $1\times 10^{-4}$ adjusted by a cosine scheduler with 8GPUs.
In the pre-training stage-2, we apply masked language modeling (MLM) \citep{devlin2018bert} upon the tokenized VQ embeddings with a 25\% random masking ratio for 500k steps. A similar pre-training setting is adopted, except the initial learning rate is $5\times 10^{-4}$ and the batch size of 2048.
In stage 3 for downstream task adaptation, we also follow the fine-tuning evaluation setting in the GUE benchmark. The pre-trained Transformer encoder is fine-tuned by AdamW with LoRA on 28 GUE datasets~\cite{Zhou2023DNABert2}, 3 EEP datasets~\cite{zhang2023gRNAdata}, and the species classification dataset~\cite{Nguyen2023HyenaDNA}.
The maximum length of the input nucleotide sequence is 512, in which case we report GFLOPs. The evaluation metrics of downstream tasks include top-1 accuracy (Acc), F1-score (F1), Matthews Correlation Coefficient (MCC), and Spearman Correlation (SC).
All experiments are implemented with PyTorch, \texttt{transformers} library, and NVIDIA A100 GPUs. The average results of 3 trials are reported. View Appendix~\ref{app:implement} and~\ref{app:downstream} for details.

\begin{table}[ht]
    \vspace{-0.5em}
    \setlength{\tabcolsep}{1.5mm}
    \centering
    \caption{Top-1 accuracy (\%) of species classification with scaling up sequence lengths, where N/A denotes out-of-memory.}
    \vspace{0.15em}
\resizebox{0.9\linewidth}{!}{
    \begin{tabular}{l|ccccc}
    \toprule
Method                & 1k         & 20k        & 32k        & 250k  & 450k  \\ \hline
HyenaDNA              & 61.13      & 87.42      & 93.42      & 97.90 & 99.40 \\
DNABERT               & 39.61      & 76.21      & 91.93      & N/A   & N/A   \\
DNABERT-2             & 61.04      & 86.83      & 99.28      & N/A   & N/A   \\
\grow\bf{VQDNA (HRQ)} & \bf{61.57} & \bf{88.05} & \bf{99.46} & N/A   & N/A   \\
    \bottomrule
    \end{tabular}
    }
    \label{tab:spec_cls}
    \vspace{-0.5em}
\end{table}

\subsection{Comparison Results}
\label{sec:comp}
We take the popular genome language models into comparison, as shown in Table~\ref{tab:overall_gue}, including DNABERT (3-mer)~\cite{Ji2021DNABert}, Nucleotide Transformer (NT) variants~\cite{Dalla2023NT}, and DNABERT-2~\cite{Zhou2023DNABert2}, where our VQDNA variants achieve the best and second best ranking of overall performances.
We first evaluate VQDNA variants on the GUE benchmark, as shown in Table~\ref{tab:gue_emp}, Table~\ref{tab:gue_pd_tfp}, and Table~\ref{tab:gue_tfp_others}, where 7 widely used genomic task are conducted, \textit{i.e.}, Epigenetic Mark Prediction (EMP) for Yeast, Transcription Factor Prediction on mouse and human genome (TFP-M and TFP-H), Covid Variants Classification (CVC), Promoter Detection (PD), Core Promoter Detection (CPD), and Splice Site Prediction (SSP). Two versions of VQDNA consistently outperform the previous large-scale model NT-2500M-1000g and the efficient model DNABERT-2 with fewer parameters, while VQDNA (HRQ) further improves VQDNA (VQVAE) by a remarkable margin.
We verify that VQDNA variants can also yield state-of-the-art performances on Editing Efficiency Prediction (EEP) with short genomic sequences in Table~\ref{tab:gue_tfp_others}.
Then, we scale up the sequence length as HyenaDNA and perform the 5-species classification task in Table~\ref{tab:spec_cls}. Although HyenaDNA can fine-tune with extremely long sequences (\textit{e.g.}, 450k), VQDNA (HRQ) achieves the best accuracy when the input sequence length is 32k (using FLASH Attention~\cite{Dao2022FlashAttention} and the gradient checkpoint technique), indicating that the learned VQDNA tokenizer can capture informative context and patterns for these extremely long-dependence tasks in genome analysis.

\begin{figure}[ht]
\vspace{-1.5em}
\begin{minipage}{0.49\linewidth}
\centering
\begin{table}[H]
    \vspace{-0.75em}
    \centering
    \setlength{\tabcolsep}{1.0mm}
    \caption{Ablation study of the total codebook size in VQDNA tokenizers.}
    \vspace{0.15em}
\resizebox{1.0\linewidth}{!}{
\begin{tabular}{c|cc|cc}
\toprule
Code size     & \multicolumn{2}{c|}{VQDNA} & \multicolumn{2}{c}{+HRQ} \\
              & Rec.         & Lin.        & Rec.        & Lin.       \\ \hline
128           & 98.2         & 42.1        & 98.4        & 42.8       \\
256           & 98.8         & 43.6        & 99.1        & 47.7       \\
\grow\bf{512} & 99.5         & \bf{44.8}   & 99.6        & \bf{48.9}  \\
1024          & \bf{99.6}    & 44.5        & \bf{99.8}   & 48.2       \\
\bottomrule
\end{tabular}
    }
    \label{tab:ab_code_size}
    \vspace{-0.75em}
\end{table}
\end{minipage}
~\begin{minipage}{0.49\linewidth}
\centering
\begin{table}[H]
    \vspace{-0.75em}
    \centering
    \setlength{\tabcolsep}{1.0mm}
    \caption{Ablation study of the codebook dimension (dim.) in VQDNA tokenizers.}
    \vspace{0.15em}
\resizebox{1.0\linewidth}{!}{
\begin{tabular}{c|cc|cc}
\toprule
Code dim.     & \multicolumn{2}{c|}{VQDNA} & \multicolumn{2}{c}{+HRQ} \\
              & Rec.      & Lin.           & Rec.     & Lin.          \\ \hline
256           & 99.4      & 44.3           & 99.5     & 48.2          \\
\grow\bf{384} & 99.5      & \bf{44.8}      & 99.6     & \bf{48.9}     \\
768           & 99.6      & 44.6           & 99.6     & \bf{48.9}     \\
1024          & 99.8      & 44.7           & 99.7     & 48.8          \\
\bottomrule
\end{tabular}
    }
    \label{tab:ab_code_dim}
    \vspace{-0.75em}
\end{table}
\end{minipage}
\vspace{-0.5em}
\end{figure}

\begin{wraptable}{r}{0.45\linewidth}
    \vspace{-4.0em}
    \setlength{\tabcolsep}{0.9mm}
    \centering
    \caption{Analysis of the mask ratio in the stage-2 MLM pre-training for our VQDNA.}
\resizebox{1.01\linewidth}{!}{
    \begin{tabular}{c|cc|cc}
    \toprule
Mask        & \multicolumn{2}{c|}{VQDNA} & \multicolumn{2}{c}{+HRQ} \\
ratio       & H3           & CVC         & H3          & CVC        \\ \hline
15\%        & 77.9         & 72.6        & 78.3        & 73.7       \\
20\%        & 78.3         & \bf{73.4}   & 78.8        & 74.2       \\
\grow{25\%} & \bf{78.6}    & 73.2        & \bf{79.2}   & \bf{74.3}  \\
30\%        & 77.4         & 73.0        & 78.6        & 73.9       \\ 
    \bottomrule
    \end{tabular}
    }
    \label{tab:ablation_mask}
    \vspace{-1.0em}
\end{wraptable}

\subsection{Ablation Study}
\label{sec:ablation}
Here, we ablate the VQ codebook settings and the mask ratio of MLM pre-training. Since applying the fine-tuning evaluation with the stage-1 tokenizers is too expensive, we report the reconstruction accuracy and the accuracy of linear probing (Lin.) ~\cite{he2021masked} on VQDNA tokenized sequences of the CVC dataset.
We first ablate the codebook dimension (dim.) and the total code size for VQDNA and HRQ. As shown in Table~\ref{tab:ab_code_size}, we found that the size of 512 is an excellent trade-off between reconstruction and discrimination abilities for both VQDNA variants, capturing more intrinsic patterns. Then, Table~\ref{tab:ab_code_dim} shows that the codebook dimension has less effect on the learned representation. Thus, we choose 384 as the default code dimension for efficiency.
Then, we analyze the masking ratio in Table~\ref{tab:ablation_mask}, reporting the fine-tuning results on the H3 and CVC datasets. We found that 25\% can help VQDNA learn better representations than 15\% or 20\% in previous models~\cite{Ji2021DNABert}. We hypothesize that VQDNA tokenizers may learn rich contextual information, allowing MLM to use large mask ratios to make the prediction task more difficult.

\begin{figure}[ht]
    \vspace{-0.25em}
    \begin{center}
    \includegraphics[width=0.875\linewidth]{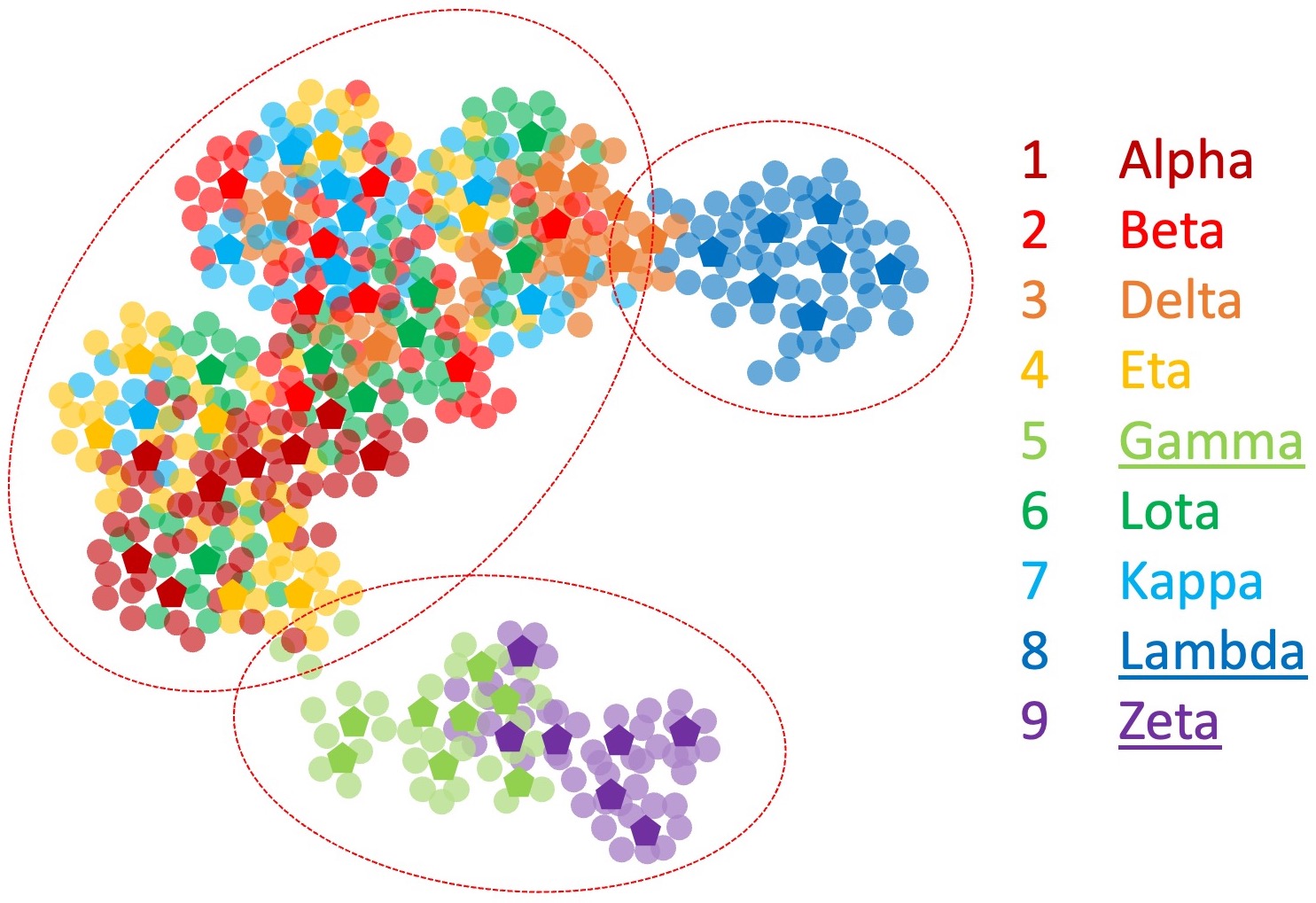}
    \end{center}
    \vspace{-1.25em}
    \caption{Visualization of the HRQ codebooks on CVC dataset by UMAP~\cite{2018UMAP}. The label of each code is obtained by calculating the most relevant class with Grad-CAM~\cite{cvpr2017gradcam} of the linear classifier learned upon HRQ-tokenized sequences. The pentagon dots stand for codes of the layer-3 codebook, while the pale circle is the layer-6 ones. The result shows great intra- \& inter-lineage pattern-awareness of HRQ vocabulary.}
    \label{fig:vq_code_vis}
    \vspace{-0.5em}
\end{figure}

\subsection{SARS-CoV-2 Analysis}
\label{sec:analysis}
SARS-CoV-2 is the cause of COVID-19, which has plunged our world into one of the gravest public health crises of the century. As the virus has proliferated globally with lightning speed, we have witnessed the rise of multiple SARS-CoV-2 variants from 2020$\sim$2021, Alpha (B.1.1.7), Beta (B.1.351), Delta (B.1.617.2), Eta (B.1.525), Lota (B.1.526), Kappa (B.1.617.1), Lambda (C.37), Gamma (P.1), Zeta (P.2), each carrying unique mutations, which are identified by Pango lineage indicators~\cite{o2022pango}.
The rapid mutation in such a short time poses an urgent and formidable challenge as it may lead to variants that could evade immune responses and resist current vaccines and treatments. 
Given the real-world significance, we conduct an empirical analysis of this issue to validate the effectiveness of the VQ tokenizer. 
Figure~\ref{fig:vq_code_vis} shows that our HRQ tokenizer learns discriminative genome embeddings, where semantically close variants (same lineage) are clustered, and the semantically distinct ones (diverse lineage) are set apart, showcasing both the \textit{intra-lineage} and \textit{inter-lineage} pattern-aware ability. Moreover, the expanded codebook successfully captures fine-grained patterns. For example, Lambda is mutated from Delta with partially similar attributes but belongs to different lineages. Lambda circles in Figure~\ref{fig:vq_code_vis} are closer to that of Delta, revealing the biological significance of HRQ. Refer to Appendix~\ref{app:covid_analysis} for detailed background and analysis.

\section{Conclusion and Discussion}
\label{sec:conclusion}
\textbf{Contributions.}
In this paper, we present VQDNA, a novel framework that leverages the VQ codebook as \textit{learnable} genome vocabulary eschewing \textit{hand-crafted} bias and rules for pattern-aware genome tokenization.
To further push the limits of the VQ tokenizer, we propose HRQ, where varying scales of codebooks are designed in a hierarchy to enrich the limited genome vocabulary in a coarse-to-fine manner.
Extensive experiments and analysis show the state-of-the-art performance of VQDNA across 32 datasets, highlighting its exceptional generalizability and biological significance.

\textbf{Limitations and Future Works.}
There are several limitations in this work:
(1) The superiority of VQDNA stems from its genome vocabulary learning, which is an additional training stage with extra costs compared to other models. Thus, there is still room for reducing its computational overhead to boost its applicability in multiple omics, as indicated by \cite{iclrw2024glms}.
(2) Due to computational constraints, the model scale of VQDNA has not reached its maximum. It is worth exploring how to scale up VQDNA with model parameters and pre-training data to increase the gained merits. For example, employing an efficient encoder with linear attention mechanisms \cite{icml2024chela, liu2024longvq} and pre-training with large-scale genomic databases \cite{Dalla2023NT, nguyen2024evo}.
(3) As the HRQ vocabulary has shown great biological significance in SARS-CoV-2 mutations, broader applications in genomics with VQDNA, such as generation tasks, deserve to be studied. Overall, all these avenues remain open for our future research.

\if\submission\VersionPublic
    \section*{Acknowledgement}
    This work was supported by the Ministry of Science and Technology of the People's Republic of China (No. 2021YFA1301603), National Natural Science Foundation of China Project (No. U21A20427), Project (No. WU2022A009) from the Center of Synthetic Biology and Integrated Bioengineering of Westlake University and Project (No. WU2023C019) from the Westlake University Industries of the Future Research Funding.
    This work was done by Zedong Wang during his research internship at Westlake University.
\fi

\section*{Impact Statement}
The goal of this paper is to advance research in multi-species genomic sequence modeling by reconstructing the tokenization to end-to-end genome vocabulary learning tasks and further introducing the pattern-aware VQDNA and HRQ to form the three-stage VQDNA training pipeline.
We have considered broader ethical impacts and do not foresee VQDNA directly leading to negative societal consequences. The genome datasets used are existing public resources that do not contain private or sensitive information. Thus, there are no privacy concerns in this study. Empirical analysis of SARS-
CoV-2 mutations reveal the biological significance of the learned
HRQ vocabulary, showcasing its potential for broader applications in genomics. We call on researchers in the community to extend this study to explore more applications with discriminative genome vocabulary learning.





\bibliography{reference}
\bibliographystyle{icml2024}


\clearpage
\renewcommand\thefigure{A\arabic{figure}}
\renewcommand\thetable{A\arabic{table}}
\setcounter{table}{0}
\setcounter{figure}{0}

\appendix


\section*{Appendix}
The appendix is structured as follows:
\begin{itemize}[leftmargin=1.25em]
    \vspace{-0.5em}
    \item In Appendix~\ref{app:implement}, we provide implementation details of training schemes of pre-training and fine-tuning stages and hyperparameter settings.
    \vspace{-0.5em}
    \item In Appendix~\ref{app:covid_analysis}, we describe background knowledge of SARS-CoV-2 variant classification and analysis.
    \vspace{-0.5em}
    \item In Appendix~\ref{app:data_pretrain} and Appendix~\ref{app:downstream}, we provide detailed information for the pre-training nucleotide database and 32 genomic downstream tasks datasets.
\end{itemize}

\section{Implementation Details}
\label{app:implement}
\paragraph{Pre-training.}
Since the two pre-training stages utilize different self-supervised methods, as shown in Figure~\ref{fig:pipeline}, we can pre-train the VQDNA tokenizer and Transformer encoder separately using human genome and multi-species genome databases (mentioned in Appendix~\ref{app:data_pretrain}).
In pre-training stage 1, the VQDNA or HRQ model is optimized by AdamW~\cite{iclr2019AdamW} ($\beta_1=0.9$, $\beta_2=0.98$, and the weight decay of 0.01), a learning rate of $1\times 10^{-4}$, and a batch size of 1024 for one epoch (around 1M steps). The hyper-parameter $\beta$ in Eq.~(\ref{eq:vqvaeloss}) and Eq.~(\ref{eq:hrqloss}) is set to 0.5 and 0.9 to balance the codebook updating and reconstruction. To stabilize training, we apply 50k steps of linear warmup with a max sequence length of 128. Then, we use the max sequence length of 256. To further improve the clustering effects of codebooks in HRQ, we employ the Repeated K-means trick~\cite{icml2023vqtorch} once after the warmup stage.
In pre-training stage 2, we store the tokenized data (as binary files) by the pre-trained VQDNA and directly load the processed nucleotide sequences to save training budges. We perform Masked Language Modeling (MLM) pre-training~\cite{Ji2021DNABert, wu2024mape} for 500k steps with a batch size of 2048, the basic learning rate of $5\times 10^{-4}$ adjusted by the cosine annealing scheduler (decay to $1\times 10^{-6}$), and a linear warmup of 10k steps. The masking ratio is 25\%, and the maximum input length is 512.

\paragraph{Fine-tuning.}
During stage 3 for downstream tasks in Figure~\ref{fig:pipeline}, the pre-trained VQDNA tokenizer and self-attention blocks in the Transformer encoder are frozen, while Low-Rank Adaptation (LoRA) machines are used for parameter-efficient fine-tuning optimized by AdamW optimizer with a batch size of 32 and a weight decay of 0.01. For each task, we choose the best combinations of the basic learning rate \{$1e-5$, $3e-5$, $5e-5$\}, the dropout rate \{0, 0.05\}, and the total fine-tuning epoch \{4, 6, 8, 10\} on the validation set, because different tasks vary in convergence difficulty and the input length. Note that the maximum input length is set to 512 during fine-tuning and uses the maximum length of each dataset during inference. We use the default LoRA hyper-parameters (a LoRA alpha is 16 and a LoRA $r$ of 8). We report the averaged results over three runs based on the optimal settings in Sec.~\ref{sec:comp}.

\section{SARS-CoV-2 Classification Analysis}
\label{app:covid_analysis}
\paragraph{Background.}
SARS-CoV-2 has continuously changed throughout the COVID-19 pandemic, resulting in multiple variants distinct from the original virus. This type of change is biologically termed a mutation-a single base change within a genome, which happens frequently but does not necessarily alter the genomic patterns of the virus. In other words, viruses with similar nucleotides may not necessarily present similar genomic patterns.
To address the concerning variants, the World Health Organization (WHO) has categorized specific viral lineages-a group of genomically related viruses descended from a common ancestor based on shared essences and properties: Variants of Interest (VOI), Variants of Concern (VOC), Variants of High Consequence (VOHC), and Variants Being Monitored (VBM). This taxonomy distinguishes well between viruses with similar nucleotides but different characteristics as an ideal classification indicator.

\paragraph{SARS-CoV-2 Variants.}
For empirical analysis, we consider the following sublineages: Alpha variants (B.1.1.7), Beta variants (B.1.351), Delta (B.1.617.2), Eta (B.1.525), Lota (B.1.526), Kappa (B.1.617.1), Lambda (C.37), Gamma (P.1), Zeta (P.2) on CVC dataset.
Based on the Pango lineage system~\cite{o2022pango}, Alpha, Beta, Delta, Eta, Lota, and Kappa are sequentially mutated, forming the Omicron (BA) lineage with similar genomic attributes. Lambda (C.37), however, is biologically mutated from Delta (B.1.617.2) which shares diverse characteristics but with more similar patterns than the others. Gamma (P.1) and Zeta (P.2) are the other two lineages, where Zeta (P.2) is mutated from Gamma (P.1) with similar genomic patterns.

\paragraph{Analysis.}
Therefore, tokenizing a genome sequence only according to its intra-sequence nucleotide bases is sub-optimal without awareness of its high-level genomic patterns. This is exactly the problem that VQ tokenizer intends to resolve. 
Ideally, the learned codebook in VQDNA can record the underlying patterns of input genomes. Specifically, different code embeddings portray different high-dimensional semantics belonging to certain groups of lineages with common attributes and characteristics, and thus, pattern-aware genome embeddings can be computed with these discriminative codebooks. Moreover, as the proposed HRQ tokenizer intends to capture more fine-grained details for hierarchical codebook learning, it is expected to distinguish the above SARS-CoV-2 variants more precisely. 
We visualize the learned codebooks by UMAP~\citep{2018UMAP}, where the labels are obtained by Grad-CAM of the linear classifier with HRQ-tokenized genome sequences. As shown in Figure~\ref{fig:vq_code_vis}, the learned codebooks are capable of distinguishing all the tested SARS-CoV-2 variants with well-aligned biological correlations. Note that the pentagon dots stand for codes of the layer-3 codebook while the circle denotes the layer-6 ones. Code embeddings belonging to each lineage (Omicron (BA), Gamma (P.1) and Zeta (P.2)) are well clustered, which indicates that they shared similar patterns for encoding different same-type variants, showcasing the exceptional \textit{intra-lineage} pattern-aware ability. More surprisingly, our HRQ codebooks can further capture the \textit{inter-lineage} patterns as the Zeta (P.2) cluster is close to the Gamma (P.1) cluster, which well exhibits their mutation correlations. Furthermore, the Delta (B.1.617.2) cluster from the Omicron lineage is near the Lambda (C.37) one, demonstrating the \textit{inter-lineage} pattern-aware capability of the HRQ codebooks.
In addition, the expanded codebook successfully captures fine-grained patterns. For example, Lambda is mutated from Delta with partially similar attributes but belongs to different lineages. Lambda circles in Figure~\ref{fig:vq_code_vis} are closer to that of Delta, which means the HRQ can capture the fine-grained patterns within genomes and lead to more precise tokenization. All this empirically confirms our claims on VQ tokenizer and reveals the biological significance of HRQ vocabulary.

\section{Multi-Species Genome for Pre-Training}
\label{app:data_pretrain}
Following~\cite{Zhou2023DNABert2}, Table \ref{tb:multi_species_details} lists the 135 species in 5 categories that we randomly selected for pre-training genome foundation models and presents the number of nucleotides collected from each species. We collected these pre-training data from the database of the National Center for Biotechnology Information (NCBI) at \url{https://www.ncbi.nlm.nih.gov/} based on Multi-species genome \url{https://huggingface.co/datasets/InstaDeepAI/multi_species_genomes} provided in Nucleotide Transformer~\cite{Dalla2023NT}.

\begin{table*}[ht]
\caption{Details statistics of the multi-species genome dataset for pre-training.}
\vspace{0.25em}
\begin{tabularx}{\textwidth}{ l p{11.3cm} c }
\toprule
\textbf{Category} & \textbf{Species} & \textbf{Nucleotides (M)} \\
\hline
Fungi & Ceratobasidium, Claviceps Maximensis, Fusarium Annulatum, Melampsora, Metschnikowia, Mucor Saturninus, Penicillium Chermesinum, Saccharomyces Cerevisiae, Sporopachydermia Quercuum, Tranzscheliella Williamsii, Xylariales & 3774 \\
\hline
Protozoa & Phytophthora Sojae, Pythium Apiculatum & 1244 \\
\hline
Mammalian & Bubalus Bubalis, Camelus Dromedarius, Human, Macaca Assamensis, 
Macaca Nigra, Mus Musculus, Peromyscus Californicus & 186931 \\
\hline
Other Vertebrate & Anas Zonorhyncha, Coregonus Clupeaformis, Gnathonemus 
Longibarbis, Myxocyprinus Asiaticus, Rhipidura Dahli & 79358 \\
\hline
Bacteria & Aeromonas, Agrobacterium, Alcaligenaceae Bacterium, Aliivibrio, Alphaproteobacteria Bacterium, Amycolatopsis Antarctica, Anaerostipes Faecis, Arthrobacter, Atopobium, Bacillus Bc15, Bacillus Bs3 2021, Bacterium, Bacteroidetes Bacterium Qs, Breoghania Corrubedonensis, Caldicoprobacter Oshimai, Candidatus Cryptobacteroides Excrementipullorum, Candidatus Dadabacteria Bacterium Rbg Combo, Candidatus Dwaynia Gallinarum, Candidatus Falkowbacteria Bacterium, Candidatus Geothermincola Secundus, Candidatus Gottesmanbacteria Bacterium, Candidatus Nomurabacteria Bacterium Full, Candidatus Portnoybacteria Bacterium Big Fil Rev, Candidatus Regiella Insecticola, Candidatus Roizmanbacteria Bacterium Combo All, Candidatus Rokubacteria Bacterium, Candidatus Saccharibacteria Bacterium, Candidatus Staskawiczbacteria Bacterium Full, Christensenella, Clostridiaceae Bacterium, Clostridiales Bacterium, Clostridium Cag 505, Clostridium Mcc328, Clostridium Nexile, Clostridium Uba3521, Collinsella Urealyticum, Coprobacillus Cateniformis, Cyanobium, Dehalococcoidia Bacterium, Enterobacteriaceae Bacterium, Evtepia Gabavorous, Firmicutes Bacterium, Fulvivirga, Jeongeupia Chitinilytica, Legionella Endosymbiont Of Polyplax Serrata, Listeria Ilorinensis, Maribacter Cobaltidurans, Marinomonas,  Mesorhizobium, Methyloceanibacter Caenitepidi, Microvirga, Mycolicibacter Engbaekii, Novosphingobium, Omnitrophica Wor Bacterium Rbg, Pantoea, Paraburkholderia Edwinii, Parerythrobacter Lutipelagi, Paulownia Witches Phytoplasma, Polaromonas Eurypsychrophila, Prevotella Ag 487 50 53, Prevotella Uba3619, Prevotella Uba634, Prochlorococcus Ag-321-I09, Prochlorococcus Ag-363-B18, Prochlorococcus Ag-402-L19, Prochlorococcus Scb243 498N4, Providencia, Pseudomonas 35 E 8, Pseudomonas Bigb0408, Pseudomonas P867, Pseudomonas Promysalinigenes, Roseobacter, Salinicola Peritrichatus, Salmonella S096 02912, Salmonella Zj-F75, Sinorhizobium, Sodalis Ligni, Sphaerochaeta, Sphingobacterium, Sphingomonas Carotinifaciens, Sphingomonas Mesophila, Sporosarcina Jiandibaonis, Sporosarcina Ureilytica, Staphylococcus Gdq20D1P, Staphylococcus M0911, Streptococcus, Streptomyces 8401, Streptomyces Di166, Streptomyces Durbertensis, Streptomyces Neau-Yj-81, Streptomyces Rk74B, Thermopetrobacter, Uncultured Kushneria, Uncultured Phascolarctobacterium, Uncultured Proteus, Verrucomicrobiales Bacterium, Vibrio, Victivallis Lenta, Virgibacillus Salexigens, Xanthomonadales Bacterium & 3610 \\
\bottomrule
\end{tabularx}
\label{tb:multi_species_details}
\end{table*}

\clearpage

\section{Downstream Datasets}
\label{app:downstream}
\subsection{GUE Benchmark}
\label{app:dataset_gue}

\begin{table*}[h]
	\centering
	\footnotesize
	\caption{
	Statistics of tasks in GUE benchmark~\cite{Zhou2023DNABert2}, including the task name, evaluation metric, and the number of training, validation, and test samples in each dataset.
	}
    \vspace{0.25em}
	\setlength{\tabcolsep}{2.4mm}{
	\begin{tabular}{lclcc}
        \toprule
\textbf{Task}                                      & \textbf{Metric} & \textbf{Datasets}            & \textbf{Train / Dev / Test} & Class \\ \hline
\multirow{3}{*}{\bf{Core Promoter Detection}}      &                 & tata                         & 4904 / 613 / 613            &       \\
                                                   & MCC             & notata                       & 42452 / 5307 / 5307         & 2     \\
                                                   &                 & all                          & 47356 / 5920 / 5920         &       \\ \hline
\multirow{3}{*}{\bf{Promoter Detection}}           &                 & tata                         & 4904 / 613 / 613            &       \\
                                                   & MCC             & notata                       & 42452 / 5307 / 5307         & 2     \\
                                                   &                 & all                          & 47356 / 5920 / 5920         &       \\ \hline
\multirow{5}{*}{\bf{Transcription Factor}}         &                 & wgEncodeEH000552             & 32378 / 1000 / 1000         &       \\
                                                   &                 & wgEncodeEH000606             & 30672 / 1000 / 1000         &       \\
                                                   & MCC             & wgEncodeEH001546             & 19000 / 1000 / 1000         & 2     \\
                                                   &                 & wgEncodeEH001776             & 27294 / 1000 / 1000         &       \\
                                                   &                 & wgEncodeEH002829             & 19000 / 1000 / 1000         &       \\ \hline
\textbf{Splice Site Prediction}                    & MCC             & reconstructed                & 36496 / 4562 / 4562         & 3     \\ \hline
\multirow{5}{*}{\bf{Transcription Factor}}         &                 & Ch12Nrf2Iggrab               & 6478 / 810 / 810            &       \\
                                                   &                 & Ch12Znf384hpa004051Iggrab    & 53952 / 6745 / 6745         &       \\
                                                   & MCC             & MelJundIggrab                & 2620 / 328 / 328            & 2     \\
                                                   &                 & MelMafkDm2p5dStd             & 1904 / 239 / 239            &       \\
                                                   &                 & MelNelfeIggrab               & 15064 / 1883 / 1883         &       \\ \hline
\multirow{10}{*}{\bf{Epigenetic Marks Prediction}} &                 & H3                           & 11971 / 1497 / 1497         &       \\
                                                   &                 & H3K14ac                      & 26438 / 3305 / 3305         &       \\
                                                   &                 & H3K36me3                     & 27904 / 3488 / 3488         &       \\
                                                   &                 & H3K4me1                      & 25341 / 3168 / 3168         &       \\
                                                   & MCC             & H3K4me2                      & 24545 / 3069 / 3069         & 2     \\
                                                   &                 & H3K4me3                      & 29439 / 3680 / 3680         &       \\
                                                   &                 & H3K79me3                     & 23069 / 2884 / 2884         &       \\
                                                   &                 & H3K9ac                       & 22224 / 2779 / 2779         &       \\
                                                   &                 & H4                           & 11679 / 1461 / 1461         &       \\
                                                   &                 & H4ac                         & 27275 / 3410 / 3410         &       \\ \hline
\textbf{Virus}                                     & F1              & Covid variant classification & 77669 / 7000 / 7000         & 9     \\
		\bottomrule
	\end{tabular}
 }
    \label{tb:gue_details}
\end{table*}

The GUE benchmark proposed by DNABERT-2 contains $28$ datasets of $7$ biological important genome analysis tasks for $4$ different species. To comprehensively evaluate the genome foundation models in modeling variable-length sequences, we select tasks with input lengths ranging from $70$ to $1000$. Table \ref{tb:gue_details} presents the detailed statistics of each evaluation dataset. The following descriptions of the supported tasks are included in the GUE benchmark~\citep{Zhou2023DNABert2}. We attach these resources here for illustration.

\paragraph{Promoter Detection (Human)} focuses on identifying (proximal) promoter regions, crucial sequences in the human genome responsible for instigating transcription. As many primary regulatory elements are located in this region, accurately detecting these sites is instrumental in advancing our grasp of gene regulation mechanisms and pinpointing the genomic underpinnings of numerous diseases. The dataset is divided twofold, TATA and non-TATA, based on whether a TATA box motif is present in the sequence. We extract -249~+50 bp around the transcription start site (TSS) from TATA and non-TATA promoters downloaded from Eukaryotic Promoter Database (EPDnew) \citep{dreos2013epdpromoter} and use it as our promoter class. Meanwhile, we construct the non-promoter class with equal-sized randomly selected sequences outside of promoter regions but with TATA motif (TATA non-promoters) or randomly substituted sequences (non-TATA, non-promoters). We also combine the TATA and non-TATA datasets to obtain a combined dataset named \textit{all}.

\paragraph{Core Promoter Detection (Human)} is similar to proximal promoter detection with a focus on predicting the core promoter region only, the central region closest to the TSS and start codon. A much shorter context window (center -34~+35 bp around TSS) is provided, making this a more challenging task than proximal promoter prediction. 

\paragraph{Transcription Factor Binding Site Prediction (Human)}  predicts binding sites of transcription factors (TF), the key proteins that regulate gene expression in the human genome. Their accurate prediction is key to deciphering complex genetic interactions and identifying potential targets for gene therapies. We accessed the legacy 690 ENCODE ChIP-seq experiments \citep{mouse2012encyclopedia} via the UCSC genome browser, encompassing 161 TF binding profiles in 91 human cell lines. We extracted a 101-bp region around the center of each peak as the TFBS class and nonoverlapping sequences with the same length and GC content as the non-TFBS class. Finally, we randomly select $5$ datasets out of a subset of 690 that we curated by heuristically filtering out tasks that are either too trivial (e.g., over 0.95 F1) or too challenging (e.g., less than 0.50 F1) for existing language models.

\paragraph{Splice Site Prediction (Human)} predicts splice donor and acceptor sites, the exact locations in the human genome where alternative splicing occurs. This prediction is crucial to understanding protein diversity and the implications of aberrant splicing in genetic disorders. The dataset \citep{wang2019splicefinder} consists of 400-bp-long sequences extracted from Ensembl GRCh38 human reference genome. As suggested by \citet{Ji2021DNABert}, existing models can achieve almost perfect performance on the original dataset, containing 10,000 splice donors, acceptors, and non-splice site sequences, which is overly optimistic about detecting non-canonical sites in reality. As such, we reconstruct the dataset by iteratively adding adversarial examples (unseen false positive predictions in the hold-out set) in order to make this task more challenging.  

\paragraph{Transcription Factor Binding Site Prediction (Mouse)} predicts the binding site of transcription factors on mouse genomes. Like human binding site data, we obtain mouse ENCODE ChIP-seq data \citep{mouse2012ChIPseq}, the largest available collection on the UCSC genome browser (n=78). This time, the negative examples are created using dinucleotide shuffling while preserving relative frequencies, while all other settings stay the same as the human TFBS prediction dataset. We also randomly select $5$ datasets out of the $78$ datasets using the same process described above.

\paragraph{Epigenetic Marks Prediction (Yeast)}  predicts epigenetic marks in yeast, modifications on the genetic material that influence gene expression without altering the DNA sequence. Precise prediction of these marks aids in elucidating the role of epigenetics in yeast. We download the $10$ datasets from \url{http://www.jaist.ac.jp/~tran/nucleosome/members.htm} and randomly split each dataset into training, validation, and test sets with a ratio of 8:1:1.

\paragraph{Covid Variant Prediction (Virus)} aims to predict the variant type of the SARS\_CoV\_2 virus based on $1000$-length genome sequences. We download the genomes from the EpiCoV database \citep{khare202SARS_CoV_2} of the Global Initiative on Sharing Avian Influenza Data (GISAID). We consider $9$ types of SARS\_CoV\_2 variants, including \textit{Alpha}, \textit{Beta}, \textit{Delta}, \textit{Eta}, \textit{Gamma}, \textit{Iota}, \textit{Kappa}, \textit{Lambda} and \textit{Zeta}.

\subsection{Additional Datasets}
\label{app:dataset_addition}
\paragraph{Editing Efficiency Prediction Dataset}
Life science studies involving clustered, regularly interspaced short palindromic repeat (CRISPR) editing generally apply the best-performing guide RNA (gRNA) for a gene of interest in human DNA. Three large-scale gRNA datasets with SpCas9/gRNA activities are provided in \citet{zhang2023gRNAdata} on K562, Jurkat, and H1 cells. The gRNA sequence is a standardized 63-length RNA encoded with ACGT, and the activity of editing efficiency is a scaler (as the regression target). The Spearman correlation is adopted as the metric to indicate high and low editing samples. The K562 dataset contains 277,000 training data and 69,262 testing data. Jurkat dataset contains 285,150 and 71,297 training and testing data. The H1 dataset has 54,580 and 13,654 training and testing samples.

\paragraph{Species Classification}
Since the discriminative mutations of different species are located in various positions, long-range dependencies are essential for discriminating different species. It requires the model to process extremely long sequences, \textit{e.g.}, up to 32k, to learn a distinct mutational profile for each species. To investigate this special issue, long-range species classification data is collected in \citet{Nguyen2023HyenaDNA}, which randomly select five species, including human (homo sapien), lemur (lemur catta), mouse (mus- culus), pig (sus scrofa), and hippo (hippopotamus amphibious). This dataset contains long genomic sequences up to 1 million lengths.



\end{document}